\documentstyle[aps,epsf,multicol]{revtex}
\newcommand{\be}{\begin{equation}}
\newcommand{\ee}{\end{equation}}
\newcommand{\ber}{\begin{eqnarray}}
\newcommand{\eer}{\end{eqnarray}}
\newcommand{\de}{\end{equation*}}
\newcommand{\cer}{\begin{eqnarray*}}
\newcommand{\der}{\end{eqnarray*}}

\draft
\begin{document}
\title{Off Resonant Pumping for Transition from Continuous to Discrete Spectrum
 and Quantum Revivals in Systems in Coherent States
 }
\author{G. S. Agarwal and J. Banerji}
\address{ Physical
Research Laboratory, Navrangpura, Ahmedabad 380 009, India}
\maketitle
\begin{abstract}
We show that in parametrically driven systems and, more generally,
in systems in coherent states, off-resonant pumping can cause a
transition from a continuum energy spectrum of the system to a
discrete one, and result in quantum revivals of the initial state.
The mechanism responsible for quantum revivals in the present case
is different from that in the non-linear wavepacket dynamics of
systems such as Rydberg atoms. We interpret the
 reported phenomena as an optical analog of
Bloch oscillations realized in Fock space and propose a feasible
scheme for inducing Bloch oscillations in trapped ions.
\end{abstract}

\pacs {{\bf Keywords:} Quantum revivals, Bloch oscillations,
optical parametric interactions}
\begin{multicols}{2}
\section{Introduction}
Parametrically driven systems have been studied extensively in
quantum and nonlinear optics. Such systems have resulted in the
production of twin beams of photons with high degrees of
entanglement\cite{ref1}. These beams of light have been used
extensively in many applications such as in connection with Bell's
inequalities and EPR paradox \cite{ref2}, quantum cryptography
\cite{ref3}, teleportation \cite{ref4}, quantum imaging
\cite{ref5} and in general, in connection with the subject of
quantum information processing \cite{ref6}. The standard
Hamiltonian describing such a parametric interaction is known to
have continuous spectrum which then results in the exponential
growth of (say) the energies in the signal and idler modes. The
state of the system at any given time is, in fact, the so-called
Perelomov coherent state \cite{perel}.

In this communication, we report unusual features in the dynamics
of systems in coherent states. We show how changing the detuning
of the field pumping the system can cause a transition from a
continuum energy spectrum of the system to a discrete one.  This
transition from continuum spectrum to discrete spectrum results in
quantum revivals of the initial state.

This study has been motivated primarily by recent work on the
optical realization of Bloch oscillations\cite{bo} and
Wannier-Stark ladders \cite{wsl}. These have been recently
observed in ultracold atoms trapped in optical potentials
\cite{boatom}, and in  fields propagating in waveguide arrays with
a linear bias so that the refractive index of a guide is
proportional to its spatial location \cite{waveguide}.
 Even the effect of nonlinearities
in the waveguide on Bloch oscillations has been studied. The
underlying mathematical equation which leads to Bloch oscillations
has the structure \be\label{e1} i\dot{C}_n =n \Delta C_n +\beta
(C_{n+1} +C_{n-1})\ee In the absence of the term $n \Delta C_n$,
an initial excitation in one of the $C$'s gets delocalized , i.e,
it spreads over all the other sites. However, if $\Delta\neq 0$,
then the system returns to the original site. The linear
dependence on the site index in the term $C_n$ is very critical
for the revival of the initial state. In the waveguide example,
the linear bias produces the linear $n$ dependence of the term $n
\Delta C_n$. The optical realization of analogs of Bloch
oscillations is based on situations where the dynamics could be
described by equations like (\ref{e1}).  We note that many
coherently driven systems have dynamics characterized by equations
like (\ref{e1}). Thus, remarkably enough, systems in coherent
states can exhibit analog of Bloch oscillations. In the case of
the radiation field, the index $n$ would refer to the occupation
number of a given Fock state. Thus the standard measurement of the
photon number distribution can be used to study the analog of
Bloch oscillations. Clearly this opens up the possibility of
studying Bloch oscillations in a very wide class of driven
systems. It is however important to have $\Delta\neq 0$, that is
that the pumping should be off-resonant. Note also that the
quantum revivals considered in this paper should be distinguished
from the well studied quantum revivals in the nonlinear wavepacket
dynamics of many systems such as Rydberg atoms \cite{rydberg} and
coherent states propagating in a Kerr medium \cite{tara}.
\section{Parametrically driven systems}
Consider a very commonly used system in optical physics. The
Hamiltonian describing the parametric interaction of two modes $a$
and $b$ with frequencies $\omega_a$ and $\omega_b$ respectively,
is given by \be\label{e2} H=\hbar (G a^\dagger b^\dagger +G^* a
b).\ee Here $G$ is the coupling constant which depends on the
nonlinearity of the medium and the amplitude of the pump. It is
assumed that the pump frequency $\omega_p=\omega_a+\omega_b$. The
operators $a$ and $b$ satisfy Bosonic commutation relations. If
the modes $a$ and $b$ are in vacuum state at time $t=0$, then the
state at time $t$ is \be\label{e3} \vert \psi(t)\rangle = \exp\{
-it(G a^\dagger b^\dagger +G^* a b)\}\vert 0,0\rangle\ee which is
the so-called Perelomov coherent state or the squeezed vacuum
state \cite{perel}. As is well known, the Heisenberg operator $a$,
for example, evolves as \be\label{e4} a(t)=a(0)\cosh (\vert G\vert
t) -i b^\dagger (0){G\over \vert G\vert} \sinh ( \vert G\vert t)
\ee and thus the energy in the mode $a$ continues to grow:
$\langle a^\dagger (t) a(t)\rangle = \sinh^2 (\vert G\vert t)$.
The probability $p_0(t)$ of finding the system in the initial
state is \ber\label{e5} p_0(t) & =&   \vert \langle 0,0\vert
\exp\{ -it(G a^\dagger b^\dagger +G^* a b)\}\vert 0,0\rangle
\vert^2\nonumber
\\
& =& \cosh^{-2}(\vert G\vert t)\sim 4 e^{-2|G| t},\quad\mbox{as
$t\to \infty$.} \eer Thus the probability of returning to the
initial state decreases monotonically and becomes exponentially
small in the long-time limit (Fig. 1). Clearly, {\it there is no
quantum revival} of the initial state.

The corresponding probability $p_n(t)$ ($n\neq 0$) of finding the
system in a state with $n$ photons in each mode, is given by
\be\label{e51} p_n(t) = \cosh^{-2}(\vert G\vert t) \tanh^{2
n}(\vert G\vert t);\quad n\neq 0.\ee Note that for $n\neq 0$,
$p_n(t)$ increases with $t$ initially and peaks at $\vert G\vert
t= \tanh^{-1}\sqrt{n/(n+1)}$. The peak value of $p_n(t)$ has the
expression \be\label{e511}
 p_n(t)\rfloor_{\rm peak}={n^n\over (n+1)^{n+1}},\ee
 which depends {\it only} on $n$ and decreases monotonically with
 $n$.
 \begin{center}
\begin{figure}[h]
\epsfxsize 8.0 cm \centerline{\epsfbox{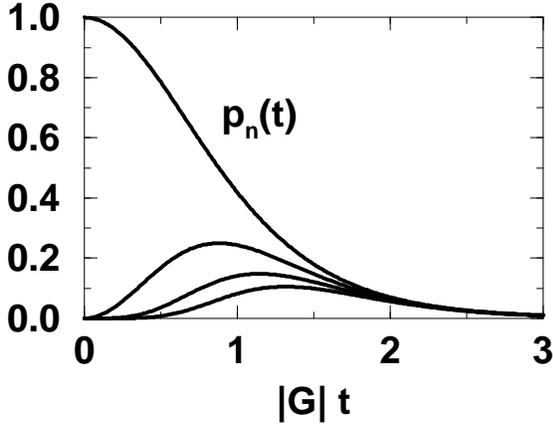}}
\caption{\narrowtext {Evolution of a parametrically driven system
when the pumping is resonant ($\Delta=0$). The system dynamics is
governed by the Hamiltonian given by Eq. (2). Here, $p_n(t)$ is
the probability that the system, starting from vacuum, reaches a
state at time $t$ with $n$ photons in each mode. Plots of $p_n(t)$
as a function of $\vert G\vert t$ are shown  for $n=0$, $1$, $2$
and $3$. The plots are identified by their peak values which
decrease as $n$ increases. Note that the probability of returning
to the initial state ($n=0$) decreases monotonically with time.}}
\end{figure}
\end{center}

The exponential growth of energies in the signal and idler modes
and the subsequent absence of quantum revival for the initial
state are connected with the fact that the Hamiltonian (\ref{e2})
 admits a continuum of
eigenvalues. The corresponding eigenstates are delta function
normalizable and signify scattering states \cite{puri}:
\ber\label{e52} H\vert\psi(\lambda)\rangle & =&
\lambda\vert\psi(\lambda)\rangle\nonumber\\
\langle\psi(\lambda')\vert\psi(\lambda)\rangle & = & \delta
(\lambda - \lambda');\quad\mbox{$\lambda$, $\lambda'$
real.}\nonumber\\ \vert\psi(\lambda)\rangle & = &  \sum_n d_n
\vert n,n\rangle , \eer where the coefficients
  $d_n$ are expressed as
hypergeometric functions of the form $_2F_1 (a,b;c,z)$.

Writing $\vert \psi(t)\rangle=\sum_n C_n(t) \vert n,n\rangle$ and
substituting in the Schrodinger equation $i\hbar
\vert\dot{\psi}\rangle=H\vert\psi\rangle$ with $H$ as given in
(\ref{e2}), we get \be\label{e6} i \dot{C}_n=G n C_{n-1}+G^* (n+1)
C_{n+1}.\ee
 Thus in terms of Fock states we have
nearest-neighbor coupling with the coupling itself depending on
the index $n$. The solutions (\ref{e5})and (\ref{e51}) imply that
with increasing time, an initial excitation in the $n=0$ state
spreads over all the states, or in the language of solid state
physics, the excitation delocalizes.

We now consider an important physical situation where the
Hamiltonian (\ref{e2}) is modified to \be\label{e7} H={\hbar
\Delta\over 2}(a^\dagger a+ b^\dagger b)+\hbar (G a^\dagger
b^\dagger +G^* a b).\ee This will be the case if the pump field is
detuned, i.e, $\omega_p\neq \omega_a+\omega_b$. The amplitudes
will satisfy Eq. (\ref{e6}) with an extra term $n\Delta C_n$ on
the right. The state at time $t$ can be obtained by using the
general disentangling theorem \cite{wod} for the SU(1,1) operators
$K_+=a^\dagger b^\dagger$, $K_-=ab$ and $K_3= (a^\dagger a+
b^\dagger b +1)/2$: \ber\label{e8}\lefteqn{
\exp\{it(GK_++G^*K_-+\Delta K_3)\}}\nonumber\\ & & =\exp(it
G\Gamma_+K_+) \exp\{(\ln \Gamma_3)K_3\}\exp(itG^*\Gamma_-K_-)\eer
where \ber\label{e9} \Gamma_3 &=& \left(\cosh \beta
-i{t\Delta\over 2 \beta}\sinh \beta\right)^{-2},\nonumber\\
 \Gamma_{\pm} & = & {2\sinh \beta\over 2 \beta \cosh \beta
 -it\Delta \sinh\beta},\nonumber\\
 \beta^2 & = & -\beta_0^2t^2,\nonumber\\
 \beta_0 & = & \sqrt{\Delta^2/4 -\vert G\vert^2}.
 \eer
 The probabilities are now given by $p_0(t)=\vert \Gamma_3\vert$ and
 $p_n(t)=\vert \Gamma_3\vert\vert\Gamma_+\vert^{2n}$
 where
 \ber\label{e10}
 \vert\Gamma_3\vert &=& \left[1+\left(\vert G\vert/\beta_0\right)^2
  \sin^2 \beta_0 t\right]^{-1},\nonumber\\ \vert \Gamma_{+}\vert^2 & = & {\sin^2
\beta_0 t \over \sin^2 \beta_0 t +(\beta_0/\vert G\vert)^2}
  \eer
  if the detuning $\Delta$ and the coupling constants are such
  that $\beta_0$ (defined by Eq. (\ref{e9})) is real. Clearly the
  survival probability of the initial state is now an oscillatory function of time $t$
  and $p_0(t)$ is unity whenever $\beta_0 t$ is an integer
  multiple of $\pi$, i.e.
  \be\label{e11}
(\Delta^2/4 -\vert G\vert^2)t^2=n^2 \pi^2,\quad\mbox{where $n$ is
an integer}. \ee
  That is,  for these times, the system returns to the original state
  $\vert 0,0\rangle$ up to an overall phase factor and there is a quantum revival (Fig. 2).
\begin{center}
\begin{figure}[h]
\epsfxsize 8.0 cm \centerline{\epsfbox{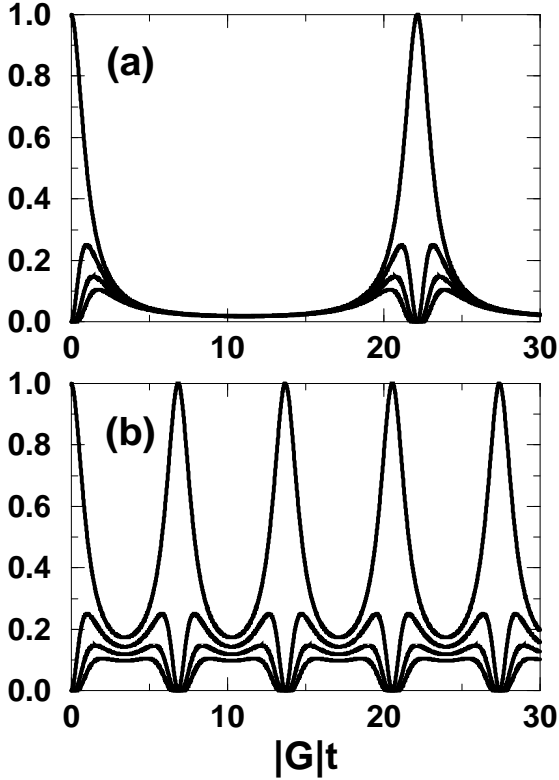}}
\caption{\narrowtext {As in Fig. 1 except that the pumping is
off-resonant ($\Delta\neq 0$). The corresponding Hamiltonian is
given by Eq. (\ref{e7}). Plots of $p_n(t)$ are shown for (a)
$\Delta/2|G|=1.01$ and (b) $\Delta/2|G|=1.10$.}}
\end{figure}
\end{center}
We thus find
  the important result that our parametric system exhibits analog
  of Bloch oscillation under the condition of non-zero detuning
  such that $\Delta >2\vert G\vert$ and that the survival
  probability becomes unity for times defined by Eq. (\ref{e11}).

  It is interesting to note that the peak values of $p_n(t)$
  ($n\neq 0$) are still given by Eq. (\ref{e511}) although their
  locations, at $\beta_0 t=\sin^{-1} (\sqrt{n} \beta_0/\vert
  G\vert)$, will now depend on the detuning $\Delta$.

   Under the condition $\Delta >2\vert G\vert$,
   the Hamiltonian (\ref{e7}) admits a discrete set of eigenvalues
  given by
  \be\label{e12} {\cal E}_{qm}=\hbar \beta_0 (q+2m+1)-\hbar
  \Delta/2;\quad\mbox{$m=0$, $1$, $2$, ...}
  \ee
where $q=a^\dagger a -b^\dagger b$ is an integer constant of
motion.
  Thus the periodic behavior is connected with a \underline{transition
  in the nature of energy levels}, i.e, the system has gone from a
  continuous spectrum to a discrete spectrum
  whose energy levels are equidistant. Note that the level separation is
  $2\hbar \beta_0$ which further validates our earlier observation that the revival period
  is given by $\pi/\sqrt{\Delta^2/4-\vert G\vert^2}$ (see Eq. (\ref{e11})).
  Comparison with
  the standard Bloch problem suggests that $\Delta$ plays the role of the
  electric field whereas the magnitude of $G$ determines the coupling to the
  nearest neighbors.
\section{Oscillator driven by a detuned  coherent field}
  All the above has analogs in the theory of coherent states for a
  {\it single} mode of the radiation field. In fact, coherent states
  will exhibit some new interesting physics if the detuning is
  nonzero. Consider the behavior of an oscillator driven by a
  detuned coherent field:
  \be\label{e13}
  H=\hbar (\Delta a^\dagger a +\epsilon a^\dagger +\epsilon^* a).
  \ee
  If the system is in a vacuum state at $t=0$,
  then the state at time $t$ is a coherent state:
  \ber\label{e14}\lefteqn{
  \vert \psi(t)\rangle=
  \exp\left[
  {i\vert\epsilon\vert^2\over \Delta^2}
  \left(\Delta t -\sin \Delta t\right)
  \right]}\nonumber\\
  && \times\left\vert -i\epsilon\left({1-e^{-i\Delta
  t}\over i\Delta}\right)\right\rangle\eer
  Consider first the case of zero detuning, $\Delta=0$. In this case, the amplitude
  $-i\epsilon t$ of the coherent state continues to increase as
  $t$ increases. The peak of the photon number distribution
  continues to move away from zero as $t$ increases and the
  probability of being in the initial state decreases
  continuously and becomes exponentially small:
  $p_0(t)=\exp(-\vert\epsilon\vert^2 t^2)$.

  For non-zero detuning, on the other hand, one obtains
  \ber\label{e15}p_n(t) & = & \vert\langle
  n\vert\psi(t)\rangle\vert^2,\quad n\geq 0;\nonumber\\
  & = & e^{-\vert\alpha\vert^2}
  {\vert\alpha\vert^{2n}\over
  n!},\quad\vert\alpha\vert=\vert\epsilon\vert t\left[{\vert\sin(\Delta
  t/2)\vert\over \Delta t/2}\right].
   \eer
   Note that the peak of $p_n(t)$ occurs at
   $n=\vert\alpha\vert^2=(2\epsilon/\Delta)^2 \sin^2(\Delta t/2)$
   which is an oscillatory function of time and vanishes whenever
   $\Delta t$ is an integer multiple of $2 \pi$. At such times,
   $p_n(t)=\delta_{n,0}$,
i.e, the system returns to the initial state $\vert 0\rangle$
(Fig. 3) up to a phase factor. Note that the period does not
depend on the strength of the driving field.
\begin{center}
\begin{figure}[h]
\epsfxsize 8.0 cm \centerline{\epsfbox{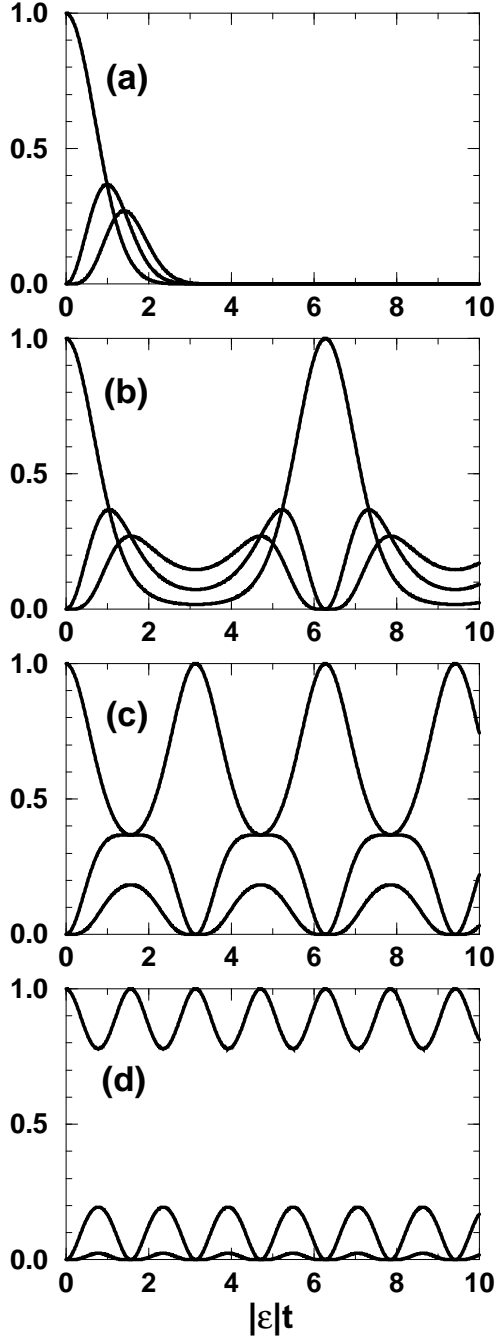}}
\caption{\narrowtext {Evolution of an oscillator driven by a
coherent field of amplitude $\epsilon$ and detuning $\Delta$. The
corresponding Hamiltonian is given by Eq. (\ref{e13}).  Plots of
$p_n(t)$ as a function of $|\epsilon| t$ are shown for $n=0$, $1$
and $2$ when (a) $\Delta/2|\epsilon|=0$, (b)
$\Delta/2|\epsilon|=0.5$, (c) $\Delta/2|\epsilon|=1.0$ and (d)
$\Delta/2|\epsilon|=2.0$. The plots are identified by their peak
values $e^{-n} n^n/n!$, which decrease as $n$ increases.  All
other notations are as in Fig. 1.}}
\end{figure}
\end{center}

 We also note that for
$\Delta\neq 0$, the eigenvalues of the Hamiltonian (\ref{e13}) are
equally spaced discrete states with energies \be\label{e151} {\cal
E}_{n}=n\hbar \Delta -\hbar\vert\epsilon\vert^2/\Delta. \ee The
spacing is independent of $\epsilon$ which leads to a periodic
behavior independent of $\epsilon$. On the other hand, if
$\Delta=0$, then $H=\hbar (\epsilon a^\dagger +\epsilon^* a)$ has
a continuum of eigenvalues with the eigenfunctions satisfying the
usual delta function condition $\langle \psi(x)\vert
\psi(x')\rangle=\delta (x-x')$. Thus here again we see the
transition from continuum eigenvalues to the discrete
set\cite{su2} and this transition is responsible for the revival
of the initial state.
\section{Quantum Revivals in ion traps}
We now show that this simple behavior can be easily probed, as for
example, for a trapped ion like the one used by Wineland and
co-workers \cite{wineland}. Consider an ion in a deep trap cooled
to its ground state. Let it interact with two Raman fields $E_1$
and $E_2$ with frequencies $\omega_1$ and $\omega_2$ such that
$\omega_1-\omega_2-\nu\neq 0$ (Fig. 4).
\begin{center} 
\begin{figure}[h]
\epsfxsize 8.0 cm \centerline{\epsfbox{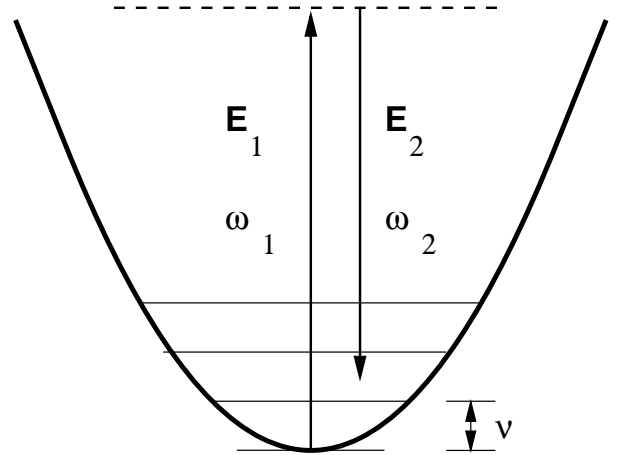}} \vspace*{0.5cm}
\caption{\narrowtext {Raman-like off-resonant transitions among
the vibrational states of an ion cooled to its electronic ground
state in a trap. The ion is acted upon by two fields with
amplitudes $E_1$ and $E_2$ and frequencies $\omega_1$ and
$\omega_2$ respectively. The separation between the relevant
vibrational levels is denoted by $\nu$ so that the detuning
$\Delta=\omega_1-\omega_2-\nu$. The ion would return to its ground
state whenever $\Delta t=2 n\pi$, where $n$ is an integer.}}
\end{figure}
\end{center}

In such a case the effective Hamiltonian describing the ionic
motion will be of the form $E_1E^*_2
a\exp\{i(\omega_1-\omega_2-\nu)t\} +h.c$. This Hamiltonian is
equivalent to (\ref {e13}). Here $a$ is the annihilation operator
for the vibrational motion of the ion.  A measurement of the
vibrational distribution will exhibit the analog of Bloch
oscillation provided $\omega_1-\omega_2-\nu\neq 0$ and the ion
would return to its ground state whenever
$(\omega_1-\omega_2-\nu)t=2\pi n$, where $n$ is an integer.
\section{Conclusion}
In conclusion, we have introduced off-resonant pumping of coherent
systems as a new mechanism for the optical analog of Bloch
oscillations. In our scheme, such oscillations occur in Fock
space. We showed that for parametrically driven systems and
systems in coherent states, the detuning of the pump changes the
 energy spectrum of the
system from continuum to discrete , and revives  the initial
state. Finally, we proposed  a scheme for inducing Bloch
oscillations in trapped ions.

\end{multicols}

\begin{references}
\bibitem{su2}{We do not consider systems belonging to
the group SU(2) as their spectrum is discrete anyway and the
system would always exhibit periodic behavior due to the
compactness of the SU(2) group}
\end{references}
\end{document}